\documentclass[aps,twocolumn,floatfix]{revtex4-1}
\usepackage{bm} 
\usepackage{dcolumn} 
\usepackage{graphicx} 
\usepackage{amsmath} 
\usepackage{amssymb} 
\usepackage{amsfonts} 

\begin{document}
\title{Plasmons in $N$-layer systems}
\author{Taehun Kim}
\author{E. H. Hwang$^{2}$}
\email{euyheon@skku.edu}
\author{Hongki Min}
\email{hmin@snu.ac.kr}
\affiliation{Department of Physics and Astronomy, Seoul National University, Seoul 08826, Korea}
\date{\today}
\affiliation{$^{2}$ SKKU Advanced Institute of Nanotechnology and Department of Nano Engineering, Sungkyunkwan University, Suwon 16419, Korea}

\begin{abstract}
In multilayer structures, the coupling between layers gives rise to unique plasmon modes, but analytic solutions are typically available only for bilayers due to the increasing complexity as the number of layers increases. We investigate plasmons in multilayer structures, including the effects of interlayer tunneling. By introducing the Coulomb eigenvector basis for multilayer systems, which can be solved exactly using Kac-Murdock-Szeg\H{o} Toeplitz matrices, we analytically derive the long-wavelength plasmon dispersions both with and without interlayer tunneling. In the $N$-layer systems, we find that, in the absence of interlayer tunneling, the out-of-phase acoustic or charge neutral plasmon modes with linear dispersions ($\omega_\alpha\propto q/\sqrt{{1-\cos{\left(\frac{\alpha-1}{N}\pi\right)}}}$ for $\alpha = 2, 3, \cdots, N$) exist, while the in-phase classical plasmon mode exhibits its conventional dispersion ($\omega_1\propto \sqrt{q}$). When interlayer tunneling is present, the out-of-phase modes develop plasmon gaps that are governed by specific interband transitions, whereas the classical mode remains unaffected. These findings have broad applicability to general coupled-layer structures.
\end{abstract}

\maketitle

{\em Introduction.} ---
Plasmons, the collective oscillations of charge carriers in a material, play a crucial role in determining the optical and electronic properties of the system \cite{Pines1966,Mahan2000,Giuliani2005}. In multilayer structures, the coupling between layers gives rise to unique plasmon modes with tunable dispersion characteristics. Recent advances in two-dimensional (2D) van der Waals materials such as graphene, transition metal dichalcogenides, and their heterostructures, have enabled the fabrication of layered systems with highly tunable electronic properties \cite{Giuliani2005, Manfra2014, Chung2021, Geim2013, Polini2020}. The ability to control interlayer interactions through stacking order \cite{Profumo2010, Jang2015, Fei2015, Choi2023}, twist angles \cite{Liu2014, Novelli2020, Zhang2020, Shin2023}, and electrostatic gating \cite{Burg2017, Nguyen2019} has facilitated the manipulation of plasmon dispersion with highly confined plasmons and strong nonlocal effects \cite{deVega2017, Alonso-Gonzalez2017, deVega2019, Sun2020, Zhu2021, Menabde2021, Goncalves2021, Chakraborty2022, Huang2022}. This necessitates a systematic approach to understanding plasmon dispersion in multilayer structures.

Conventional random phase approximation (RPA) methods are typically applied in a layer basis to obtain plasmon mode dispersions. However, due to the increasing complexity of the Hamiltonian as the number of layers increases, analytical solutions are generally limited to bilayer systems \cite{DasSarma1981, DasSarma1998, Borghi2009, Hwang2009, DasSarma2009, Gamayun2011, Roldan2013, Hwang2018, Mohammadi2021, Afanasiev2022}, while numerical approaches are typically employed for multilayer systems \cite{Jang2015, Chakraborty2022, Zhu2013, Wachsmuth2014, Gumbs2016, Lin2018, Kim2020, VanMen2021, Cavicchi2024}. In this paper, we investigate plasmons in multilayer structures, incorporating the effects of interlayer tunneling. To obtain analytic results, we introduce the Coulomb eigenvector basis, which can be solved exactly using Kac-Murdock-Szeg\H{o} (KMS) Toeplitz matrices \cite{Kac1953, Trench2001, Bogoya2016, Fikioris2019, Narayan2021}. This approach allows for an analytic determination of long-wavelength plasmon dispersions and applies to systems both with and without interlayer tunneling. 
\begin{figure}[htb]
\includegraphics[width=1.0\linewidth]{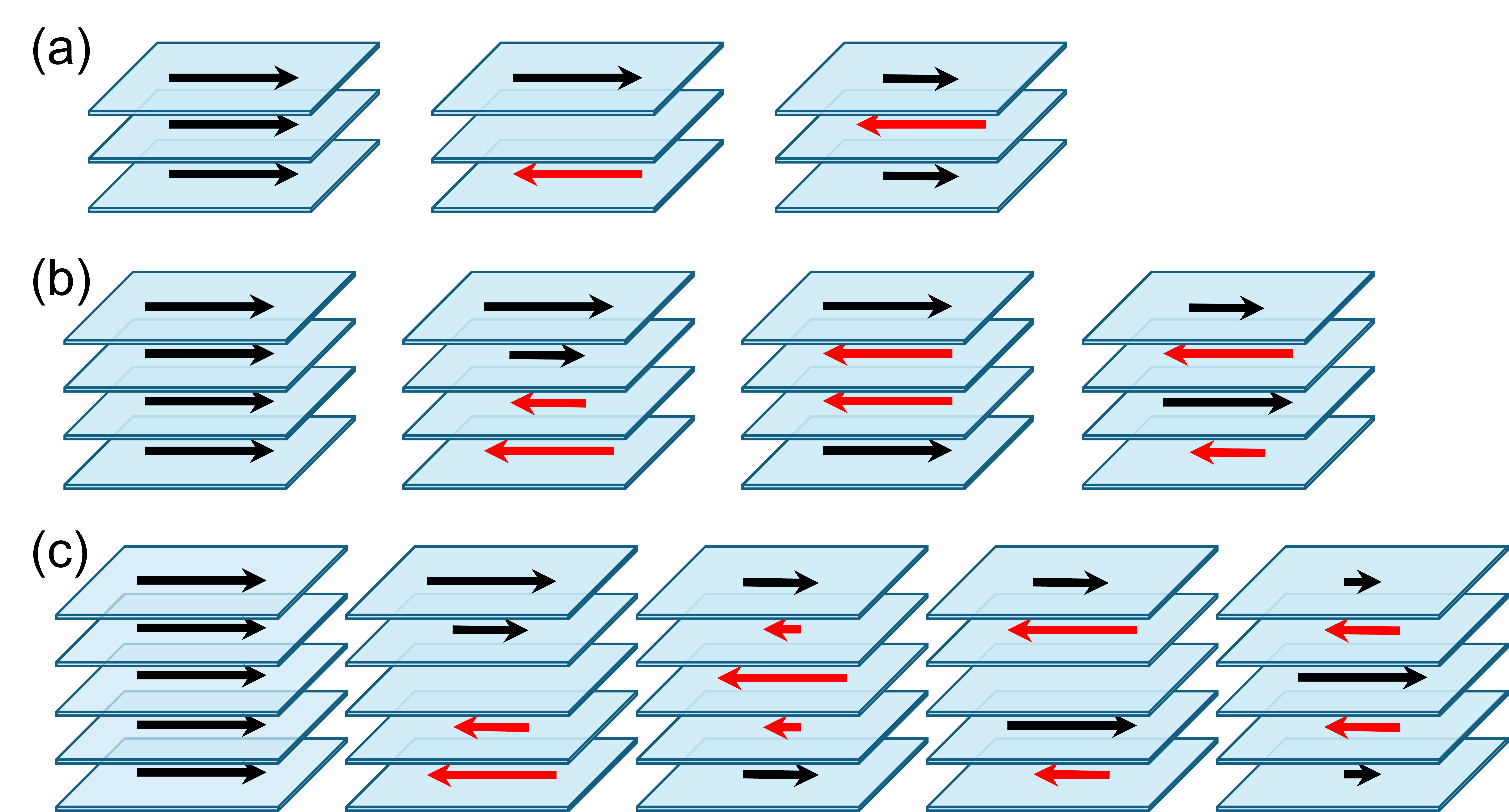}
\caption{
Schematic illustrations of the Coulomb eigenvectors in the long-wavelength limit for (a) a trilayer, (b) a tetralayer, and (c) a pentalayer system. The first column shows the in-phase mode, whereas the remaining columns depict the charge-neutral out-of-phase modes.
} 
\label{fig:fig1}
\end{figure}
We find that for decoupled $N$-layer systems, $N-1$ out-of-phase acoustic or charge neutral plasmon modes exist with linear dispersions ($\omega_{\alpha}=v_{\alpha} q$ where $v_{\alpha}\propto 1/\sqrt{1 - \cos\left(\frac{\alpha - 1}{N}\pi\right)}$ for $\alpha=2,3,\cdots,N$) and one in-phase classical mode with a square-root dispersion ($\omega_1\propto \sqrt{q}$). 
When interlayer tunneling is present, we find that the out-of-phase plasmon modes develop plasmon gaps that are governed by specific interband transitions ($\omega_{\alpha}=\sqrt{\omega_{\text{gap}, \alpha}^2 + C_{\alpha} q}$ for $\alpha=2,3,\cdots,N$), whereas the in-phase mode $\omega_1$ remains qualitatively unaffected by interlayer tunneling.
We note that the $N-1$ out-of-phase modes in multilayer systems are charge-neutral plasmon modes (see Fig.~\ref{fig:fig1}), which are related to Pines' demon mode (i.e., a charge-neutral collective excitation arising from a zero net charge oscillation between two different bands in three-dimensional (3D) systems \cite{Pines1956,Husain2023,Zhao2023}). However, unlike the demon mode, these out-of-phase modes in multilayer systems are tunable via interband transitions.

{\em Kac-Murdock-Szeg\H{o} matrix.} ---
Consider a single-particle Hamiltonian $H(\boldsymbol{k})$ for an $N$-layer system with an interlayer separation $d$. Assume that the corresponding eigenstates are given by $|\boldsymbol{k}, \lambda\rangle$ with energy dispersions $\varepsilon_{\boldsymbol{k}, \lambda}$ and a band index $\lambda$.
The noninteracting density-density response function in the layer basis is given by \cite{Mahan2000,Giuliani2005}
\begin{equation} \label{eq:noninteracting_response-layer_basis}
\begin{split}
    \chi_{ij}^{(0)}(\boldsymbol{q}, \omega) =& g\sum_{\lambda,\lambda'} \int {d^2k\over(2\pi)^2} \frac{f_{\boldsymbol{k}, \lambda}-f_{\boldsymbol{k+\boldsymbol{q}}, \lambda'}}{\omega + \varepsilon_{\boldsymbol{k}, \lambda} - \varepsilon_{\boldsymbol{k+\boldsymbol{q}}, \lambda'}+i\eta}\\
    &\times F_{ij}^{\lambda\lambda'}(\boldsymbol{k}, \boldsymbol{k+\boldsymbol{q}}),
\end{split}
\end{equation}
where $i, j=1,2,\cdots,N$ are the layer indices, $g$ is the spin-valley degeneracy factor, and $F_{ij }^{\lambda\lambda'}(\boldsymbol{k},\boldsymbol{k}')=\langle\boldsymbol{k}, \lambda|P_i|\boldsymbol{k}', \lambda'\rangle \langle \boldsymbol{k}', \lambda'|P_j|\boldsymbol{k}, \lambda\rangle$ is the wavefunction overlap factor with the projection operator $P_{i}$ onto the $i$th layer, $f_{\boldsymbol{k}, \lambda}= [e^{\beta(\varepsilon_{\boldsymbol{k}, \lambda}-\mu)}+1]^{-1}$ is the Fermi-Dirac distribution at chemical potential $\mu$, and $\eta$ is a positive infinitesimal. (Here, we set $\hbar=1$.)

Within the RPA, the matrix element of the dielectric function in the layer basis can be obtained as
\begin{equation}\label{eq:dielectric_function-layer_basis}
    \begin{split}
        \epsilon_{ij}(\boldsymbol{q}, \omega) = \delta_{ij} - \sum_{k}V_{ik}(q)\chi_{kj}^{(0)}(\boldsymbol{q}, \omega),
    \end{split}
\end{equation}
where $V_{ij}(q) = v(q)e^{-|i-j|qd}$ is the Coulomb matrix describing the interaction strength between layers separated by a distance of $|i-j|d$ with the 2D Coulomb potential $v(q) = \frac{2\pi e^2}{\kappa q}$ and a background dielectric constant $\kappa$. Since obtaining the plasmon dispersions using Eq.~({\ref{eq:dielectric_function-layer_basis}}) for $N>2$ is, in general, extremely difficult, we introduce a new approach based on the Coulomb eigenvector basis.
It is important to note that the Coulomb matrix is a special case of a Toeplitz matrix known as the KMS matrix, which is defined by $A_{ij}(\rho)=\rho^{|i-j|}$ for $0 < \rho < 1$. Thus, the Coulomb matrix can be expressed as $V_{ij}(q)=v(q) A_{ij}(\rho)$ with $\rho=e^{-qd}$. The corresponding eigenvalues $g_{\alpha}(\rho)$ and eigenfunctions $\boldsymbol{u}_\alpha(\rho)=\frac{1}{N_\alpha}(u_{\alpha}^{(1)}(\rho), u_{\alpha}^{(2)}(\rho), \cdots, u_{\alpha}^{(N)}(\rho))^T$ of the KMS matrix $A_{ij}(\rho)$ with normalization constant $N_\alpha$ ($\alpha=1,2,\cdots,N$) are given by [see Sec.~\ref{KMS_MATRIX} of the Supplemental Material (SM) \cite{Supplemental} and references \cite{Kac1953, Trench2001, Bogoya2016, Fikioris2019, Narayan2021}]
\begin{subequations}\label{eq:KMS_eigenvalue_eigenvector}
\begin{flalign}
        \label{eq:KMS_eigenvalue}
        g_{\alpha}(\rho)&=\frac{1-\rho^2}{1-2\rho\cos(\theta_{\alpha})+\rho^2},\\
        \label{eq:KMS_eigenvector}
        u_{\alpha}^{(k)}(\rho) &= \sin\left[\frac{\alpha}{N+1}k\pi + \left(\frac{1}{2} - \frac{k}{N+1}\right)\eta_{\alpha}(\rho)\right],
\end{flalign}
\end{subequations}
where $\eta_{\alpha}(\rho)=2\arctan\left[{\rho\sin(\theta_{\alpha})\over 1-\cos(\theta_{\alpha})}\right]$ and $\theta_{\alpha}$ ($\frac{\alpha-1}{N}\pi<\theta_\alpha<\frac{\alpha}{N}\pi$) is the unique solution of the equation 
\begin{equation}\label{eq:eta_equation}
    (N+1) \theta_{\alpha} + \eta_{\alpha}(\rho) = \alpha \pi.
\end{equation}

In the Coulomb eigenvector basis, the dielectric function can be expressed as
\begin{equation}
    \begin{split}
        \epsilon_{\alpha\beta}(\boldsymbol{q}, \omega) = \delta_{\alpha\beta} - V_\alpha(q)\chi^{(0)}_{\alpha\beta}(\boldsymbol{q}, \omega),
    \end{split}
    \label{dielectric_Coulomb}
\end{equation}
where $V_{\alpha}(q)=v(q)g_{\alpha}(e^{-qd})$ represents the eigenvalues of the Coulomb matrix. Here, $\chi^{(0)}_{\alpha\beta}(\boldsymbol{q}, \omega)$ is the noninteracting density-density response function in the Coulomb eigenvector basis, which can be obtained by replacing the overlap factor $F_{ij}^{\lambda\lambda'}(\bm{k}, \bm{k'})$ in Eq.~(\ref{eq:noninteracting_response-layer_basis}) with that in the Coulomb eigenvector basis, $F_{\alpha\beta}^{\lambda\lambda'}(\bm{k}, \bm{k'}) = \langle\bm{k}, \lambda|U_{\alpha}|\bm{k'}, \lambda'\rangle \langle\bm{k'}, \lambda'|U_\beta|\bm{k}, \lambda\rangle$ where $U_\alpha = \text{diag}[\bm{u}_{\alpha}(e^{-qd})]$.
Note that odd-indexed Coulomb eigenvectors exhibit symmetry with respect to the midpoint $k = N/2$, while even-indexed Coulomb eigenvectors are characterized by antisymmetry, as shown in Fig.~\ref{fig:fig1}.
Unlike the layer basis, which requires full matrix diagonalization, the Coulomb eigenbasis provides a strong selection rule for interband transitions, allowing for a more tractable analytical treatment. Specifically, when a system consists of bands that are either symmetric or antisymmetric, interband transitions between symmetric and antisymmetric bands are allowed only through antisymmetric Coulomb modes, while other transitions occur via symmetric Coulomb modes.

{\em Plasmons in multilayer structures.} ---
We consider multiple quantum wells with an effective mass $m$, interlayer tunneling $t$, and a spatial separation $d$, which is an extension of the two coupled quantum well system \cite{DasSarma1998}. 
The Hamiltonian matrix is given by $H_{ij}(\boldsymbol{k}) = k^2/2m$ for $i=j$, $t$ for $|i-j|=1$, and 0 otherwise. The energy levels are $\varepsilon_{\boldsymbol{k}, \lambda} =  k^2/2m + \Delta_{\lambda}$ with $\Delta_\lambda = -2t\cos(\phi_\lambda)$ and the corresponding wavefunctions are given by $|\lambda\rangle = \sqrt{\frac{2}{N+1}}(\psi^{(1)}_{\lambda}, \psi^{(2)}_{\lambda}, \cdots, \psi^{(N)}_{\lambda})^{T}$, where $\psi^{(k)}_{\lambda} = (-1)^k\sin{(k\phi_\lambda)}$ and $\phi_\lambda = \frac{ \lambda}{N+1}\pi$ with $\lambda = 1, 2, \cdots, N$, which are the usual solutions to the one-dimensional chain problem. Note that the wavefunctions can be categorized as either symmetric or antisymmetric, similar to the Coulomb eigenvectors in Eq.~(\ref{eq:KMS_eigenvector}).

In the absence of tunneling ($t = 0$), we can write the matrix $\chi^{(0)}$ in the layer basis as $\chi_{ij}^{(0)}=\chi^{(0)}_{\rm{2D}}\delta_{ij}$, where $\chi^{(0)}_{\rm{2D}}$ is the noninteracting response function of a single-layer 2D electron gas. In this case, the eigenvectors of the Coulomb interaction in Eq. (\ref{eq:KMS_eigenvector}) correspond to the normal modes of the decoupled multilayer system since they diagonalize the dielectric function.
Thus, it is straightforward to obtain the plasmon modes by solving $\text{det}[\epsilon(\boldsymbol{q}, \omega)] = 0$. 
In the long-wavelength limit, we have the following low-energy behavior (see Sec.~\ref{sec:derivation_decoupled} of the SM \cite{Supplemental}):
\begin{subequations}\label{eq:absence of tunneling}
\begin{flalign}
        \omega_1^2(q \rightarrow 0) &= \frac{2\pi e^2 n_{\text{tot}}}{\kappa m} q, \\
        \omega_{\alpha\neq 1}^2(q \rightarrow 0) &= \frac{2\pi e^2 d n_{\text{tot}}}{\kappa m N\left[1 - \cos\left(\frac{\alpha - 1}{N}\pi\right)\right]} q^2 \label{gapless out-of-phase dispersion},
\end{flalign}
\end{subequations}
where $n_{\text{tot}}$ is the total 2D electron density. This shows that a decoupled $N$-layer system naturally features the well-known classical mode ($\omega \propto \sqrt{q}$) along with $N-1$ acoustic modes ($\omega \propto q$). The dispersion of the classical mode is determined by $n_{\text{tot}}$, while the acoustic modes are governed by the average electron density per layer ($n_{\text{tot}} / N$), with their velocities depending on $\alpha$. 

In the presence of tunneling ($t \neq 0$), the non-diagonal elements of $\chi^{(0)}$ must be considered. In the long-wavelength limit, $\chi^{(0)}$ in the Coulomb eigenvector basis takes the form (see Sec. \ref{Calculation of the density-density response} of the SM \cite{Supplemental})
\begin{equation}\label{long-wavelength limit susceptibility}
        \text{Re}[\chi^{(0)}_{\alpha\beta}(\boldsymbol{q}, \omega)] = \sum_{\lambda,\lambda'} F^{\lambda\lambda'}_{\alpha\beta} n_{\lambda'} \frac{2\left(\Delta_{\lambda\lambda'} + \frac{q^2}{2m}\right)}{\omega^2 - \left(\Delta_{\lambda\lambda'}+\frac{q^2}{2m}\right)^2},
\end{equation}
where $n_\lambda$ is the electron density in the $\lambda$ band and $\Delta_{\lambda\lambda'} = \Delta_\lambda - \Delta_{\lambda'}$ is the interband splitting. Note that the overlap factor $F_{\alpha\beta}^{\lambda\lambda'}$ in the Coulomb eigenvector basis ensures that elements of $\chi^{(0)}_{\alpha\beta}(\bm{q}, \omega)$ remain nonzero only when both $\alpha$ and $\beta$ are either even or odd. Furthermore, for the $\alpha$th Coulomb oscillation, it can be shown that for given $\lambda$ and $\lambda'$, the overlap factor indicates that certain terms become dominant when $|\lambda-\lambda'|=\alpha-1$, $\lambda+\lambda'=\alpha-1$, or $\lambda +\lambda' = 2(N+1) - (\alpha - 1)$, leading to only specific interband transitions contributing to the oscillation. These transitions correspond to out-of-plane momentum transfers in the folded Brillouin zone, as will be discussed in the $N\rightarrow \infty$ limit later. 
See Sec. \ref{Calculation of the density-density response} of the SM \cite{Supplemental} for details. By calculating the dominant interband contributions, we find one gapless in-phase mode and $N-1$ gapped out-of-phase modes given by (see Sec.~\ref{sec:derivation_coupled-strong_weak} of the SM \cite{Supplemental})
\begin{subequations}
    \begin{flalign}
    \omega_{1}^2(q \to 0) &= \frac{2\pi e^2 n_{\text{tot}}}{\kappa m} q,\\
    \omega_{\alpha \neq 1}^{2}(q \to 0) &= \omega_{\text{gap}, \alpha}^2 + C_{\alpha} q, 
    \end{flalign}
\end{subequations}
where $\omega_{\text{gap},\alpha}=\omega_\alpha(q=0)$ is the plasmon gap of the out-of-phase mode for $\alpha\neq 1$, and $C_{\alpha}$ is a nonzero constant when $t\neq 0$, which can be obtained from a Taylor expansion of $g_{\alpha}(e^{-qd})$ and $u_{\alpha}(e^{-qd})$ in powers of $q$. 
It is important to note that the in-phase mode remains unaffected by interlayer tunneling, depending only on the total electron density $n_{\text{tot}}$, while the out-of-phase modes develop a gap due to the effects of interlayer tunneling, generalizing the coupled bilayer case. We now focus on the plasmon gaps.

In the strong tunneling regime ($t/\varepsilon_{\rm F} \gg q_{\rm{TF}}d$ where $q_{\rm{TF}} = 2me^2 / \kappa$ is the 2D Thomas-Fermi wave vector and $\varepsilon_{\rm F}$ is the Fermi energy), 
plasmon gaps are governed by specific interband transitions. If the density is confined to the lowest energy band ($n_1=n_{\text{tot}}$), we obtain $\omega_{\text{gap}, \alpha}$ as
\begin{eqnarray}\label{eq:plasmon_gap-strong_tunneling}
\omega_{\text{gap}, \alpha}^2 &=& \Delta_{\alpha1}^2 + \frac{2q_{\text{TF}}d\pi}{m} \\
&\times&\left[\frac{F_{\alpha\alpha}^{\alpha1}}{1-\cos\left(\frac{\alpha-1}{N}\pi\right)}+\frac{F_{\alpha+2, \alpha+2}^{\alpha1}}{1-\cos\left(\frac{\alpha+1}{N}\pi\right)}\right]n_1\Delta_{\alpha1}. \nonumber
\end{eqnarray}
Here, we take $F_{\alpha\alpha}^{\lambda\lambda'}=0$ when $\alpha>N$. It is evident that the plasmon gaps exhibit the behavior $\omega_{\text{gap}, \alpha}\approx \Delta_{\alpha1}$ in the low-density limit. Note that each out-of-phase mode is determined by a single interband transition, while one or two Coulomb oscillations $\bm{u}_{\alpha}$ and $\bm{u}_{\alpha+2}$ contribute to each out-of-phase plasmon mode. Calculating the overlap factors also enables the identification of the most dominant Coulomb oscillation.

\begin{figure}[htb]
\includegraphics[width=0.985\linewidth]{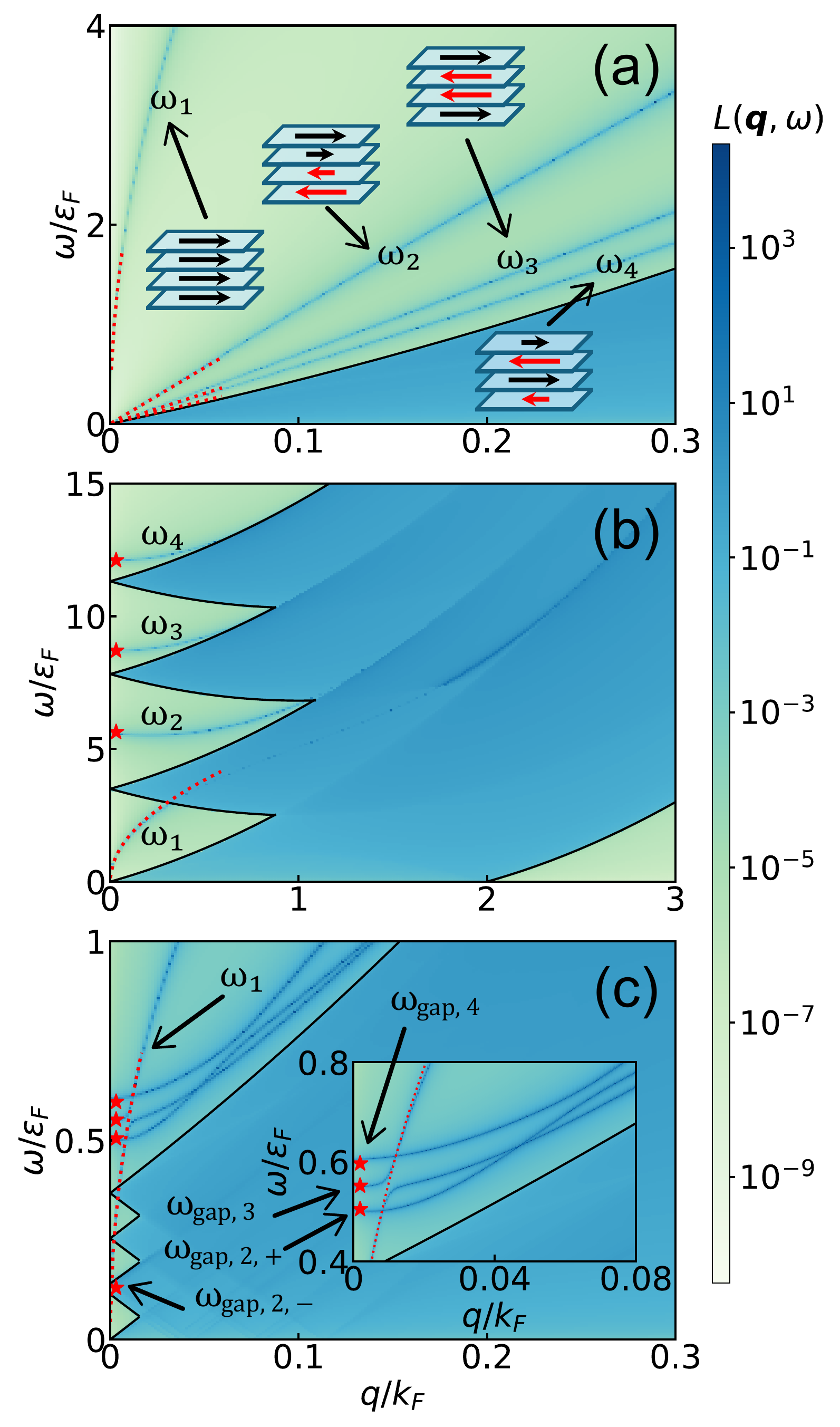}
\caption{Loss functions $L(\boldsymbol{q}, \omega) = -\text{Im}\left\{\text{Tr}\left[\epsilon^{-1}(\boldsymbol{q}, \omega)\right]\right\}$ of a tetralayer system for (a) $n_{\text{tot}} = 4\times 10^9 \text{ cm}^{-2}$ in the absence of tunneling, (b) $n_{\text{tot}} = 4\times 10^9 \text{ cm}^{-2}$ in the presence of tunneling, and (c) $n_{\text{tot}} = 4\times 10^{11} \text{ cm}^{-2}$ in the presence of tunneling, respectively. The thick black lines denote the boundaries of the single particle excitations. In (a), Coulomb oscillations for each plasmon mode are illustrated with out-of-phase modes having zero net charge oscillations. The inset in (c) highlights the long-wavelength plasmon dispersions. The red dotted lines represent the calculated analytical plasmon dispersions, and the red stars indicate the analytically obtained plasmonic gaps. For the calculations, the parameters corresponding to a GaAs quantum well are used: $m = 0.067m_{\rm e}$, $\kappa=10.9$, $t=0.5$ meV, and $d=200$ {\AA} with $\eta = 5\times10^{-5}\varepsilon_{\rm F}$, where $m_{\rm e}$ is the electron mass.
}
\label{N=4_loss}
\end{figure}

In the weak tunneling regime ($t/\varepsilon_{\rm F} \ll q_{\rm{TF}}d$), multiple bands become occupied, which means various types of interband transitions typically appear for different Coulomb modes. Nevertheless, we only need to consider the diagonal components of $\chi^{(0)}_{\alpha\beta}$ as in the case of no tunneling, since the off-diagonal terms are sufficiently small compared to the diagonal terms. By solving $1-V_{\alpha}\chi^{(0)}_{\alpha\alpha}=0$, the plasmon gap for each mode $\alpha$ can be obtained, which is primarily governed by a single Coulomb oscillation. To leading order in tunneling, the plasmon gap approximately has the form
\begin{equation}\label{eq:plasmon_gap-weak_tunneling}
\omega_{\text{gap}, \alpha}\propto \sqrt{{q_{\text{TF}}d\pi\Delta_{\alpha1}\over m}}.
\end{equation}

Note that for odd $N$, there are $(N+1)/2$ symmetric Coulomb eigenmodes including one in-phase mode, and $(N-1)/2$ antisymmetric Coulomb eigenmodes. 
Similarly, for even $N$, there are $N/2$ symmetric Coulomb modes including one in-phase mode and $N/2$ antisymmetric modes. 
In all cases, the symmetric and antisymmetric modes remain decoupled, with plasmon gaps arising from interband transitions.


Here, we show a tetralayer system as a specific example (see Sec.~\ref{sec:derivation_N3} \cite{Supplemental} of the SM for a trilayer system). In the long-wavelength limit, we have two symmetric oscillation modes [$\boldsymbol{u}_{1}\parallel (1, 1, 1, 1)^T$ and $ \boldsymbol{u}_3 \parallel (1, -1, -1, 1)^T$] and two antisymmetric oscillation modes [$\boldsymbol{u}_2 \parallel (1+\sqrt{2}, \sqrt{2}, -\sqrt{2}, -1-\sqrt{2})^T$ and $\boldsymbol{u}_4  \parallel (\sqrt{2}-1, -1, 1, 1-\sqrt{2})^T$]. In the absence of tunneling [see Fig.~\ref{N=4_loss}(a)], the long-wavelength plasmon modes can be obtained through simple calculations using Eq. (\ref{eq:absence of tunneling}).
In the strong tunneling limit [see Fig.~\ref{N=4_loss}(b)], when only the lowest band is occupied ($n_1\neq0$ and $n_2=n_3=n_4=0$), from Eq.~(\ref{eq:plasmon_gap-strong_tunneling}), we obtain
\begin{subequations}\label{eq:plasmon_gap-strong_tunneling-N4}
\begin{flalign}
\omega_{\text{gap}, 2}^2 &= \Delta_{21}^2 + \frac{q_{\text{TF}}d\pi}{m}\frac{6n_1}{5}\Delta_{21}, \\
\omega_{\text{gap}, 3}^2 &= \Delta_{31}^2 + \frac{q_{\text{TF}}d\pi}{m}\frac{2 n_1}{5}\Delta_{31}, \\
\omega_{\text{gap}, 4}^2 &= \Delta_{41}^2 + \frac{q_{\text{TF}}d\pi}{m}\frac{(\sqrt{5}-1+\sqrt{2})^2}{20}n_1\Delta_{41}.
\end{flalign}
\end{subequations}
In the weak tunneling limit [see Fig.~\ref{N=4_loss}(c)], the off-diagonal terms can be ignored, allowing us to focus solely on the diagonal terms.
Note that in the tetralayer case, the energy spacing between adjacent bands differs, leading to a mixture of different interband transitions. As a result, the out-of-phase mode $\omega_2$ exhibits distinct behavior compared to the other modes. From Eq.~(\ref{long-wavelength limit susceptibility}), $\omega_2$ is determined as a solution to the quadratic equation in $\omega^2$, implying that it always has two solutions: a symmetric solution ($\omega_{2+}$) and an antisymmetric solution ($\omega_{2-}$). 
This phenomenon closely resembles the case of a conventional double-well potential problem, where energy levels split into symmetric and antisymmetric states.
Finally, we obtain
\begin{subequations}\label{eq:plasmon_gap-weak_tunneling-N4}
\begin{flalign}
\!\! \omega_{\text{gap}, 2+}^2 &=\frac{\Delta_{21}^2 + \Delta_{32}^2}{2} + C, \\
\!\! \omega_{\text{gap}, 2-}^2 &= \frac{\Delta_{21}^2+\Delta_{32}^2}{2}, \\
\!\! \omega_{\text{gap}, 3}^2 &= \Delta_{31}^2 + \frac{q_{\text{TF}}d\pi}{m}\frac{2(n_1-n_3+n_2-n_4)}{5}\Delta_{31}, \\
\!\! \omega_{\text{gap}, 4}^2 &= \Delta_{41}^2 + \frac{q_{\text{TF}}d\pi}{m}\frac{(\sqrt{5}-1+\sqrt{2})^2(n_1-n_4)}{20}\Delta_{41}.
\end{flalign}
\end{subequations}
See Sec.~\ref{sec:derivation_N4} of the SM \cite{Supplemental} for the detailed derivation and the definition of $C$. Note that the $\omega_{2-}$ mode remains near the interband transition region. 
For the third mode $\omega_{3}$, which is symmetric, only symmetric-symmetric or antisymmetric-antisymmetric bands contribute to the oscillation, while for the fourth mode $\omega_{4}$, which is antisymmetric, only symmetric-antisymmetric bands contribute to the oscillation.
Since both the in-phase mode $\omega_1$ and the third mode $\omega_3$ are generated by symmetric Coulomb modes, the $\sqrt{q}$ dispersion of the in-phase mode $\omega_1$ couples with the other symmetric out-of-phase mode $\omega_3$ in the weak interlayer tunneling regime, but it cannot couple with the antisymmetric out-of-phase modes, $\omega_2$ and $\omega_4$, as shown in the inset to Fig.~\ref{N=4_loss}(c).

{\em Plasmons in the bulk limit.} ---
For an $N\rightarrow \infty$ system, the bulk plasmon dispersions can be obtained by treating $\theta_\alpha$ as a continuous variable and replacing $\theta_\alpha \rightarrow q_z d$ in Eq.~(\ref{eq:eta_equation}).  In the absence of tunneling, the plasmon modes are given by (see Sec.~\ref{sec:derivation_N_infinite} of the SM \cite{Supplemental})
\begin{equation}\label{eq:plasmon-infinite_N}
    \omega_{q_z}^2(q\rightarrow 0) = \frac{2\pi e^2n_{\text{2D}}}{\kappa m}\frac{\sinh(qd)}{\cosh(qd) - \cos(q_zd)}q,
\end{equation}
where $n_{\rm{2D}}$ is the electron density of a single layer. This result is consistent with the well-known infinite-layer Fetter model \cite{Fetter1974, Jain1985, Jain1985a, Christian2022}. From Eq.~(\ref{eq:plasmon-infinite_N}), it follows that $q_z$ corresponds to the out-of-plane wavevector. 
In the presence of tunneling, we consider $\Delta_{\lambda}$ as $\Delta_{k_z} = -2t\cos(k_zd)$ by replacing $\frac{\lambda}{N+1}\pi \rightarrow k_z d$. Under this transformation, the overlap factor can be interpreted as a delta function, and the dominant interband transition occurs when $|k_z-k_z'|=q_z$, $k_z+k_z'=q_z$, or $k_z + k_z' = 2\pi - q_z$. If $k_z$ is extended from $-\pi/d$ to $\pi/d$, the band index summation is exactly replaced by an integration over the Brillouin zone for the out-of-plane wavevector. Notably, the dominant transition can be reformulated as occurring between $k_z$ and $k_z+q_z$, leading to the following expression for the noninteracting density-density response function:
\begin{equation}
\!\! \frac{\chi^{(0)}_{q_z}(\boldsymbol{q}, \omega)}{d}\! = g \!\int\frac{d^3k}{(2\pi)^3}\!\frac{f_{\bm{k}, k_z}-f_{\bm{k+q}, k_z+q_z}}{\omega+\varepsilon_{\bm{k}, k_z} - \varepsilon_{\bm{k+q}, k_z+q_z}+i\eta},
\end{equation}
where $\varepsilon_{\bm{k}, k_z} = \varepsilon_{\bm{k}} + \Delta_{k_z}$ and $f_{\bm{k}, k_z} = [e^{\beta(\varepsilon_{_{\bm{k}, k_z}}-\mu)}+1]^{-1}$, which represents the conversion of the band index into a continuous variable $k_z$. This result is identical to the noninteracting density-density response function in a 3D anisotropic system. Specifically, for $q_z d\ll 1$, by substituting $\Delta_{q_z}\rightarrow q_z^2/2m_z$, the plasmon dispersion in the long-wavelength limit is given by \cite{Ahn2021}
\begin{equation}
    \omega_{q_z}^2(q\rightarrow 0) = \frac{4\pi e^2n_{\text{3D}}}{\kappa m_z} + \bigg(\frac{q_z^2}{2m_z}\bigg)^2,
\end{equation}
where $n_{\text{3D}}$ denotes the electron density in a 3D electron system.

{\em Conclusion.} ---
In summary, we have systematically investigated plasmons in multilayer structures, incorporating the effects of interlayer tunneling. By employing the Coulomb eigenvector basis and using the Kac-Murdock-Szeg\H{o} Toeplitz matrices, we have derived analytical solutions for long-wavelength plasmon modes in both coupled and uncoupled systems. We find that in the absence of interlayer tunneling, $N-1$ out-of-phase acoustic plasmon modes show linear dispersions $\omega_{\alpha}= v_{\alpha} q$ with the slope $v_{\alpha}=\sqrt{\frac{2\pi e^2 d n_{\text{tot}}}{\kappa m N\left[1 - \cos\left(\frac{\alpha - 1}{N}\pi\right)\right]}}$ ($\alpha=2,3,\cdots,N$), which are charge-neutral excitations similar to Pines' demon mode, and one in-phase classical mode with a square-root dispersion $\omega_1=\sqrt{\frac{2\pi e^2 n_{\text{tot}}}{\kappa m} q}$. When interlayer tunneling is introduced, the $N-1$ out-of-phase plasmon modes develop plasmon gaps governed by specific interband transitions, whereas the in-phase mode remains unaffected in both cases. 
Importantly, our analytic results for plasmons in multilayer systems are broadly applicable to van der Waals heterostructures and other layered materials, offering potential applications in optoelectronics and plasmonic device engineering.

%
%
%
%

\acknowledgments
The work at SNU was supported by the National Research Foundation of Korea (NRF) grants funded by the Korea government (MSIT) (Grant No. RS-2023-NR076715), the Creative-Pioneering Researchers Program through Seoul National University (SNU), and the Center for Theoretical Physics.
EHH acknowledges support from the National Research Foundation of Korea (NRF) (Grant No. RS-2021-NR058646).

\section*{data availability}
The data are available upon reasonable request from the authors.


\clearpage
\widetext
\setcounter{section}{0}
\setcounter{equation}{0}
\setcounter{figure}{0}
\setcounter{table}{0}
\renewcommand{\theequation}{S\arabic{equation}}
\renewcommand\thefigure{S\arabic{figure}} 
\setcounter{page}{1}
\begin{center}
\textbf{\large Supplemental Material for ``Plasmons in $N$-layer systems"}
\end{center}
\begin{center}
{Taehun Kim,$^{1}$ E. H. Hwang,$^{2\ast}$ and Hongki Min$^{1\dagger}$}\\
\emph{$^{1}$ Department of Physics and Astronomy, Seoul National University, Seoul 08826, Korea}\\
\emph{$^{2}$ SKKU Advanced Institute of Nanotechnology and Department of Nano Engineering, Sungkyunkwan University, Suwon 16419, Korea}
\end{center}

\section{Kac-Murdock-Szeg\H{o} Toeplitz matrices} \label{KMS_MATRIX}

In this section, we review the properties of the Kac-Murdock-Szeg\H{o} (KMS) matrix $A_{ij} = \rho^{|i-j|}$ for $0 < \rho < 1$ and $1\leq i,j \leq N$ \cite{SM_Kac1953, SM_Trench2001, SM_Bogoya2016, SM_Fikioris2019, SM_Narayan2021}: 
\begin{equation}
    A= \begin{bmatrix}
         1 & \rho & \rho^2 & \cdots & \rho^{N-2} & \rho^{N-1}\\
        \rho & 1 & \rho & \cdots & \rho^{N-3} & \rho^{N-2}\\
        \rho^2 & \rho & 1 & \cdots & \rho^{N-4} & \rho^{N-3} \\
        \vdots & \vdots & \vdots & \ddots & \vdots & \vdots \\
        \rho^{N-2} & \rho^{N-3} & \rho^{N-4} & \cdots & 1 & \rho \\
        \rho^{N-1} & \rho^{N-2} & \rho^{N-3} & \cdots  & \rho & 1
    \end{bmatrix}.
\end{equation}

In order to obtain the eigenvalues $g$ and eigenvectors $\bm{u} = (u_1, u_2, \cdots, u_N)^T$ of $A$, consider its inverse $B=A^{-1}$ given by the following tridiagonal matrix:
\begin{equation}
    B= \frac{1}{1-\rho^2}\begin{bmatrix}
         1 & -\rho & 0 & \cdots & 0 & 0\\
        -\rho & 1+\rho^2 & -\rho & \cdots & 0 & 0\\
        0 & -\rho & 1+\rho^2 & \cdots & 0 & 0 \\
        \vdots & \vdots & \vdots & \ddots & \vdots & \vdots \\
        0 & 0 & 0 & \cdots & 1+\rho^2 & -\rho \\
        0 & 0 & 0 & \cdots  & -\rho & 1
    \end{bmatrix}.
\end{equation}
Note that the eigenvectors $\bm{u}$ of $A$ also diagonalize $B$, with corresponding eigenvalues given by $g^{-1}$. Then, we have the following equations
\begin{equation}\label{eq:eigenvector_relations}
    -\rho u_{k-1} 
    + \left( 1 + \rho^2 - \frac{1-\rho^2}{g} \right) u_k  
    - \rho u_{k+1} 
    = 0, \quad 1 \leq k \leq N,
\end{equation}
with the boundary conditions
\begin{equation}
    u_0 = \rho u_1 \quad \text{and} \quad u_{N+1} = \rho u_{N},
\end{equation}
where we take $u_0$ and $u_{N+1}$ for convenience. Then, the components of the eigenvectors are of the form $u_k = c_1 z^k + c_2 z^{-k}$ where $z$ and $z^{-1}$ are the zeros of the polynomial
\begin{equation}\label{eq:reciprocal polynomial}
     -\rho z^2 + \left(1+\rho^2 - \frac{1-\rho^2}{g}\right)z - \rho = 0.
\end{equation}
From the boundary conditions, we obtain the equation $C\begin{pmatrix} c_1 \\ c_2 \end{pmatrix} =0$, where
\begin{equation}
    C =
    \begin{bmatrix}
        1-\rho z && 1-\rho z^{-1} \\
        z^{N+1}-\rho z^{N} && z^{-N-1} - \rho z^{-N}
    \end{bmatrix},
\end{equation}
which implies that, for the nontrivial solutions, we need to solve the determinant 
\begin{equation}\label{eq:determinantC_N}
    \det{C} = (z^{-N-1} - z^{N+1}) - 2\rho (z^{-N} - z^{N}) + \rho^2(z^{-N+1} - z^{N-1}) = 0.
\end{equation}
With the given Eqs.~(\ref{eq:reciprocal polynomial}) and (\ref{eq:determinantC_N}), we always have a pair of solutions $z$ and $z^{-1}$ satisfying
\begin{equation}
    g = \frac{1 - \rho^2}{1-\rho (z + z^{-1}) + \rho^2}.
\end{equation}
Since the matrix $A$ is symmetric and has real eigenvalues and eigenvectors, by setting $z=e^{i\theta}$, we can determine the eigenvalues and eigenvectors. From Eq.~(\ref{eq:determinantC_N}), we define
\begin{equation}
    D_{N}(\theta) = \frac{\sin{(N+1)\theta}}{\sin{\theta}} - 2\rho \frac{\sin{N\theta}}{\sin{\theta}} + \rho^2 \frac{\sin{(N-1)\theta}}{\sin{\theta}},
\end{equation}
where $\theta\in (0, \pi)$. Note that $c_1, c_2$ being nontrivial is a necessary and sufficient condition for $D_N(\theta)=0$. One can see that
\begin{subequations}\label{eq:Change signs}
\begin{flalign}
    D_N(\theta \rightarrow0) &= (1-\rho)[N(1-\rho) + 1+\rho], \\
    D_N(\theta=\frac{\alpha\pi}{N}) &=(-1)^\alpha (1-\rho^2), \quad \alpha=1, 2, \cdots, N-1, \\
    D_N(\theta\rightarrow \pi) &= (-1)^N(1+\rho)[N(1+\rho)+1-\rho].
\end{flalign}
\end{subequations}
From Eq.~(\ref{eq:Change signs}), $D_N(\theta)$ changes sign in each interval $(\frac{\alpha-1}{N}\pi, \frac{\alpha}{N}\pi)$ for $\alpha = 1, 2, \cdots, N$, which implies that $D_N(\theta)$ has one solution $\theta_\alpha$ in $(\frac{\alpha-1}{N}\pi, \frac{\alpha}{N}\pi)$ since $A$ has $N$ eigenvalues. Substituting $(N+1)\theta_\alpha = \alpha\pi - \eta_{\alpha}(\rho)$ into $D_N(\theta)$, $\eta_\alpha(\rho)$ can be obtained by
\begin{eqnarray}
    \eta_\alpha(\rho) &= &\arctan{\left[\frac{2(1-\rho\cos{\theta_\alpha})\rho\sin{\theta_\alpha}}{(1-\rho\cos{\theta_\alpha})^2-\rho^2\sin^2{\theta_\alpha}}\right]} \nonumber \\
    &=& 2\arctan{\left[\frac{\rho\sin{\theta_\alpha}}{1-\rho\cos{\theta_\alpha}}\right]}, 
\end{eqnarray}
which corresponds to Eq.~(\ref{eq:eta_equation}). Here, we use the identity $2\arctan{x} = \arctan{\frac{2x}{1-x^2}}$. Thus, $g_{\alpha}(\rho)$ and $\bm{u}_{\alpha}(\rho)$ in Eq.~(\ref{eq:KMS_eigenvalue_eigenvector}) are eigenvalues and eigenvectors of $A$. 

It is important to note that in the limit $\rho\rightarrow 1$, we have $\theta_\alpha\rightarrow \frac{\alpha-1}{N}\pi$, and the corresponding eigenvalues and eigenvectors become
\begin{subequations}\label{eq:rho->1 limit}
    \begin{flalign}
        g_\alpha(\rho\rightarrow1) &= \begin{cases}
        N,&  \alpha=1, \\ 
        \frac{1-\rho}{1-\cos(\frac{\alpha-1}{N}\pi)}, & 2\leq\alpha\leq N,
        \end{cases}\\
        u_{\alpha}^{(k)}(\rho\rightarrow 1) &= \cos\left[\frac{(2k-1)(\alpha-1)}{2N}\pi\right].
    \end{flalign}
\end{subequations}

\section{Derivation of the plasmon dispersions for decoupled $N$-layer systems}\label{sec:derivation_decoupled}

The noninteracting density-density response function of a single-layer 2D electron gas $\chi^{(0)}_{\rm{2D}}(\bm{q}, \omega)$ at zero temperature can be expressed as 
\begin{eqnarray}
    \chi^{(0)}_{\rm{2D}}(\bm{q}, \omega) 
    &=&    g \int {d^2k\over(2\pi)^2} \frac{f_{\boldsymbol{k}}-f_{\boldsymbol{k}+\boldsymbol{q}}}{\omega + \varepsilon_{\boldsymbol{k}} - \varepsilon_{\boldsymbol{k+\boldsymbol{q}}}+i\eta} \nonumber \\
    &=& \frac{g m }{2\pi}\frac{k_{\rm{F}}}{q}\left[ \Psi_2 \left( \frac{\omega^+}{qv_{\rm{F}}} - \frac{q}{2k_{\rm{F}}} 
 \right) - \Psi_2 \left( \frac{\omega^+}{qv_{\rm{F}}} + \frac{q}{2k_{\rm{F}}} \right) \right] ,
\end{eqnarray}
where $\varepsilon_{\boldsymbol{k}}=k^2/2m$, $\omega^+ = \omega+i\eta$ with a positive infinitesimal $\eta$, $k_{\rm{F}}$ is the Fermi wave vector, and $v_{\rm{F}} = k_{\rm{F}} / m$ is the corresponding Fermi velocity. Here, $\Psi_{2}(z)$ is a complex function defined by \cite{SM_Giuliani2005}
\begin{equation}
    \Psi_{2}(z) = \int_0^1 dx x\int_0^{2\pi}\frac{d\phi}{2\pi} \frac{1}{z - x\cos{\theta}} = z - {\rm sign}({\rm Re}[z])\sqrt{z^2 - 1},
\end{equation}
where ${\rm sign}(x) = 1, 0, -1$ for $x>0, x=0,$ and $x<0$, respectively. Notice that $\Psi_{2}(z)$ is antisymmetric ($\Psi_2(z) = -\Psi_2(-z)$) and the leading terms of the expansion in powers of $z^{-1}$ for $|z|\rightarrow\infty$ are $\Psi_2(z)\rightarrow 1/2z$.

In the absence of tunneling, the noninteracting density-density response function in the layer basis is given by $\chi^{(0)}_{ij} = \chi^{(0)}_{\rm{2D}}\delta_{ij}$. Then, the dielectric function in the layer basis can be written as
\begin{equation}
    \epsilon_{ij}(\bm{q}, \omega) = \delta_{ij} - V_{ij}(q)\chi^{(0)}_{\rm{2D}}(\bm{q}, \omega),
\end{equation}
which implies that we only need to diagonalize the Coulomb matrix to obtain the collective excitations. By transforming into the Coulomb eigenvector basis, the plasmon modes can be obtained by solving
\begin{equation}\label{eq:determinat in the absence of tunneling}
    \det{[\epsilon(\bm{q}, \omega)]} = \prod_{\alpha =1}^N \left[1 - V_\alpha(q)\chi^{(0)}_{\rm{2D}}(\bm{q}, \omega)\right] = 0.
\end{equation}
By using Eq.~(\ref{eq:rho->1 limit}) and the asymptotic behavior of $\Psi_2(z)$, 
\begin{eqnarray}
{\rm Re}\left[1 - V_\alpha(q)\chi^{(0)}_{\rm{2D}}(q\rightarrow 0, \omega)\right] &\approx& 1 - V_{\alpha}(q) \frac{gm}{2\pi}\frac{k_{\rm{F}}}{q}\left[\frac{qv_{\rm{F}}}{2\omega - q^2/m} - \frac{qv_{\rm{F}}}{2\omega + q^2/m}\right] \nonumber \\
&\approx& 1 - V_{\alpha}(q)\frac{n_{\rm{2D}}}{m}\frac{q^2}{\omega^2} = 0,
\end{eqnarray}
 where we use $n_{\rm{2D}} = gk_{\rm{F}}^2 / 4\pi$. 
 
 For a general case, one can consider the energy dispersion $\varepsilon_{\bm{k}} = \alpha k^J$ of each decoupled layer. 
We assume $\omega\ll 2\varepsilon_{\rm F}$, such that interband effects become negligible due to Pauli blocking in the long-wavelength limit. Then, the noninteracting response function takes the form
\begin{eqnarray}
    \chi^{(0)}_{\rm{2D}}(\bm{q} \rightarrow 0, \omega) &=& g \int \frac{d^2 k}{(2\pi)^2} \frac{f_{\bm{k}} - f_{\bm{k+q}}}{\omega + \varepsilon_{\bm{k}} - \varepsilon_{\bm{k+q}} + i\eta} \nonumber \\
    &=& \frac{g}{4\pi^2}\int_{0}^{k_{\rm{F}}}k dk \int_{0}^{2\pi} d\theta \left[\frac{1}{\omega + \varepsilon_{\bm{k}} - \varepsilon_{\bm{k+q}} + i\eta} - \frac{1}{\omega - \varepsilon_{\bm{k}} + \varepsilon_{\bm{k}+\bm{q}} + i\eta}\right] \nonumber  \\
    &=&- \frac{g}{2\pi^2 \omega}\int_0^{k_{\rm{F}}} kdk \int_0^{2\pi} d\theta \left[\frac{\varepsilon_{\bm{k}}-\varepsilon_{\textbf{k+q}}}{\omega}  + \cdots\right]  \nonumber \\
    &\approx& \frac{g}{2\pi^2 \omega^2}\int_0^{k_{\rm{F}}} kdk\int_0^{2\pi} d\theta \alpha\left[J k^{J-1}q \cos{\theta} + \frac{J}{2}k^{J-2}q^2(1 + (J-2)\cos^2{\theta}) + \cdots \right]  \nonumber  \\
    &\approx& \frac{n_{\rm{2D}}}{m_{\rm F}}\frac{q^2}{\omega^2}.
\end{eqnarray}
Here, we use the relation $\varepsilon_{\bm{k}+\bm{q}} = \alpha (k^2 + 2kq \cos{\theta} + q^2)^{J/2}$ and define $m_{\rm F}=k_{\rm{F}} / v_{\rm{F}} = k_{\rm{F}}^{2-J}/J\alpha$ with $v_{\rm{F}} = J\alpha k^{J-1}_{\rm{F}}$. Note that the electron density remains $n_{\rm{2D}} = g k_{\rm{F}}^2/4\pi$ for the energy dispersion $\varepsilon_{\bm{k}} = \alpha k^J$. Thus, we obtain the long-wavelength plasmon dispersions in Eq.~(\ref{eq:absence of tunneling}) with $m$ replaced by $m_{\rm F}$.

\section{Calculation of the noninteracting density-density response function}\label{Calculation of the density-density response}

The noninteracting density-density response function of multiple quantum wells can be analytically calculated at zero temperature. The expression for $\chi^{(0)}_{\alpha\beta}(\boldsymbol{q}, \omega)$ is given as follows:
\begin{eqnarray}\label{eq:density-density response in presence of tunneling}
        \chi^{(0)}_{\alpha\beta}(\boldsymbol{q}, \omega) &=& \frac{gm}{2\pi} \sum_{\lambda,\lambda'}  \frac{F_{\alpha\beta}^{\lambda\lambda'}k_{\text{F}, \lambda '}}{q}\Theta(\mu-\Delta_{\lambda'})\Bigg[  \Psi_2\left(\frac{\omega_{\lambda'\lambda}^+}{v_{\text{F}, \lambda'} q} - \frac{q}{2k_{\text{F}, \lambda'}}\right)- \Psi_2\left(\frac{\omega_{\lambda\lambda'}^+}{v_{\text{F}, \lambda'} q} + \frac{q}{2k_{\text{F}, \lambda'}}\right) \Bigg],
\end{eqnarray}
where $\Theta(x)$ is a step function, $\omega_{\lambda\lambda'}^+ = \omega + \Delta_{\lambda\lambda'} +  i\eta$ describes the interband transition, $k_{\text{F}, \lambda}$ is the Fermi wave vector for band $\lambda$, and $v_{\text{F}, \lambda} = \frac{ k_{\text{F}, \lambda}}{m}$ is the corresponding Fermi velocity. This indicates that $\chi^{(0)}_{\alpha\beta}(\boldsymbol{q}, \omega)$ can be calculated in a manner analogous to the single-layer case by substituting $\omega \rightarrow \omega_{\lambda\lambda'}$. 

In the long-wavelength limit, Eq.~(\ref{long-wavelength limit susceptibility}) can be derived from Eq.~(\ref{eq:density-density response in presence of tunneling}) by using the asymptotic form of $\Psi_2(z)$ and considering the overlap factor of the $N$-layer system given by
\begin{eqnarray} \label{eq: U_alpha relation}
\langle\lambda|U_\alpha|\lambda'\rangle 
        &=& \sum_{k=1}^{N}\frac{\sin\left(\frac{\lambda k\pi}{N+1}\right)\sin\left(\frac{\lambda'k\pi}{N+1}\right)\cos\left[\frac{(2k-1)(\alpha-1)}{2N}\pi\right]}{{(N+1)\sqrt{(2-\delta_{\alpha, 1})N}} } \nonumber \\
        &=& \left[1 - (-1)^{\lambda-\lambda' + \alpha}\right]\frac{S(x_+, y) - S(x_-, y)}{{(N+1)\sqrt{(2-\delta_{\alpha, 1})N}} },
\end{eqnarray}
where we define $x_\pm = \frac{\lambda \pm \lambda'}{2N+2}\pi$, $y = \frac{\alpha - 1}{2N}\pi$, and $S(x, y) = \frac{\cos(y)\sin^2(x)}{\sin(y+x)\sin(y-x)}$ for simplicity. This formulation highlights that the dominant contribution occurs when $\lambda-\lambda'+\alpha$ is odd and the denominator of $S(x_+, y)$ or $S(x_-, y)$ approaches zero. In other words, the condition is satisfied when $y+x_{\pm} \approx k\pi$ or $y-x_{\pm} \approx k\pi$ for some integer $k$, which corresponds to $|\lambda'-\lambda| = \alpha - 1$, $\lambda + \lambda' = \alpha - 1$, or $\lambda+\lambda'=2(N+1) - (\alpha -1)$. 

Furthermore, this overlap factor behaves like a Dirac-delta function. For example, if $N\gg 1$, we have
\begin{eqnarray} \label{eq: delta function relation}
    \frac{S(x_+, y)^2}{N}\ &=& \frac{\cos^2(y)\sin^4\left(\frac{N}{N+1}y + \frac{\delta\alpha\pi}{N+1}\right)}{N\sin^2\left(\frac{2N+1}{N+1}y + \frac{\delta\alpha\pi}{N+1}\right)\sin^2\left(\frac{1}{N+1}y - \frac{\delta\alpha\pi}{N+1}\right)} \nonumber \\
    &=&\frac{\cos^2\left[(N+1)\epsilon\right]\sin^4(N\epsilon)}{N\sin^2\left[(2N+1)\epsilon\right]\sin^2(\epsilon)} \nonumber \\
    &\approx& \frac{\pi}{4}\delta(\epsilon),
\end{eqnarray}
where $\delta\alpha = \frac{\alpha-1 - \lambda-\lambda'}{2}$ and $\epsilon = \frac{1}{N+1}y - \frac{\delta\alpha\pi}{N+1}$ are defined for simplicity. Here, we use $\displaystyle\lim_{N\rightarrow\infty}\sin(N\epsilon)/\sin(\epsilon)\approx\displaystyle\lim_{N\rightarrow\infty}\sin(N\epsilon) / \epsilon = \pi\delta(\epsilon)$, which confirms that only a few dominant terms contribute to the density-density response function. Note that the $1 - (-1)^{\lambda-\lambda'+\alpha}$ constraint ensures that the matrix is block-diagonalized as $\chi^{(0)} = \chi^{(0)}_{\text{odd}} \oplus \chi^{(0)}_{\text{even}}$, since only elements of $\chi^{(0)}_{\alpha\beta}$ where both $\alpha$ and $\beta$ are either even or odd remain nonzero.
This implies that the oscillating modes are decoupled into symmetric and antisymmetric components. Our analysis of the overlap factor demonstrates that, even in the strong tunneling limit, only the diagonal and adjacent diagonal elements within each block are significant, while all other components are negligible.

\section{Derivation of the plasmon dispersions for coupled $N$-layer systems in the strong and weak tunneling regimes}\label{sec:derivation_coupled-strong_weak}

In the strong tunneling limit, for the out-of-phase mode $\omega_\alpha$ with $\alpha>1$, if only the lowest energy band is occupied ($n_1\neq0$ and $n_2=n_3=\cdots=0$), we can approximate the noninteracting density-density response function in the long-wavelength limit as
\begin{eqnarray}
    \chi^{(0)}_{\alpha\beta}(q\rightarrow 0, \omega) &=& \sum_{\lambda = 1}^N F_{\alpha\beta}^{\lambda 1} n_1 \frac{2\Delta_{\lambda 1} }{\omega^2 -\Delta_{\lambda1}^2} \nonumber \\
    &\approx &\begin{cases}
        F_{\alpha\alpha}^{\alpha 1}n_1 \frac{2\Delta_{\alpha 1}}{\omega^2-\Delta_{\alpha 1}^2} + F_{\alpha\alpha}^{\alpha-2, 1} n_1\frac{2\Delta_{\alpha-2, 1}}{\omega^2 - \Delta_{\alpha-2, 1}^2}, &\quad \rm{\beta = \alpha}, \\
        F_{\alpha, \alpha+2}^{\alpha 1}n_1 \frac{2\Delta_{\alpha 1}}{\omega^2-\Delta_{\alpha 1}^2}, &\quad \beta = \alpha+2, \\
        0, &\quad \rm{otherwise,}
    \end{cases}
\end{eqnarray}
where we select the dominant contributions of the $\chi^{(0)}_{\alpha\beta}$ from Eqs.~(\ref{eq: U_alpha relation}) and (\ref{eq: delta function relation}). In this case, the determinant equation can be approximated as follows when $\omega$ is near $\Delta_{\alpha 1}$:
\begin{eqnarray}\label{eq:w_alpha_strong_coupling_limit}
    \det[{\epsilon}(q\rightarrow 0, \omega)] &\approx&   
        \left[1-V_\alpha\chi^{(0)}_{\alpha\alpha}\right]\left[1-V_{\alpha+2}\chi^{(0)}_{\alpha+2, \alpha+2}\right] - V_{\alpha}V_{\alpha+2}\left[\chi^{(0)}_{\alpha, \alpha+2}\right]^2 \nonumber \\
        &\approx& 1 - (V_{\alpha}F_{\alpha\alpha}^{\alpha 1} + V_{\alpha+2}F_{\alpha+2, \alpha+2}^{\alpha 1})\frac{2n_1 \Delta_{\alpha 1}}{\omega^2 - \Delta_{\alpha 1}^2} = 0.
\end{eqnarray}
Here, we use the relation $F_{\alpha\alpha}^{\lambda\lambda'}F_{\beta\beta}^{\lambda\lambda'} = (F_{\alpha\beta}^{\lambda\lambda'})^2$ and consider only the terms that include $1/(\omega^2 - \Delta_{\alpha 1}^2)$ since $V_{\beta}n_1 \Delta_{\beta 1} /(\omega^2 - \Delta_{\beta 1}^2) \propto \varepsilon_{\rm F} q_{\rm{TF}} d / t \ll 1$ with $\beta \neq \alpha$. This clearly shows that the out-of-phase mode $\omega_{\alpha}$ is determined by the dominant interband transition $\Delta_{\alpha 1}$, and the plasmon gap becomes
\begin{equation}\label{eq:SM_plasmon gap in the strong-coupling limit}
\omega_{\text{gap}, \alpha}^2 = \Delta_{\alpha1}^2 + 2(V_{\alpha}F_{\alpha\alpha}^{\alpha1}+V_{\alpha+2}F_{\alpha+2, \alpha+2}^{\alpha1})n_1\Delta_{\alpha1},
\end{equation}
leading to Eq.~(\ref{eq:plasmon_gap-strong_tunneling}) in the main text.

In the weak tunneling regime, the off-diagonal terms are sufficiently small, as they vanish exactly in the absence of tunneling. This indicates that the matrix can be effectively approximated as diagonal, consistent with the results for decoupled systems. Consequently, the plasmon modes can be determined by solving $1-V_{\alpha}\chi^{(0)}_{\alpha\alpha}=0$ for each $\alpha$, and the noninteracting density-density response function in the long-wavelength limit is given by
\begin{eqnarray}\label{eq:noninteracting_response_weak_tunneling}
\chi^{(0)}_{\alpha\alpha}(q\rightarrow 0, \omega) &\approx\sum\limits_{\langle\lambda,\lambda'\rangle } F_{\alpha\alpha}^{\lambda\lambda'} n_{\lambda'} \frac{2\left(\Delta_{\lambda,\lambda'} + \frac{q^2}{2m}\right)}{\omega^2 - \left(\Delta_{\lambda\lambda'} + \frac{q^2}{2m}\right)^2},
\end{eqnarray}
where $\langle \lambda, \lambda'\rangle$ denotes the dominant interband transitions determined from the overlap factor relations. 
For the out-of-phase modes, these transitions are approximately replaced by $\Delta_{\alpha 1}$, and Eq.~(\ref{eq:noninteracting_response_weak_tunneling}) can be further approximated as
\begin{eqnarray}
\chi^{(0)}_{\alpha\alpha}(q\rightarrow 0, \omega) &\approx \sum\limits_{{\substack{\langle \lambda, \lambda'\rangle, \\\lambda > \lambda' } }}F_{\alpha\alpha}^{\lambda\lambda'}(n_{\lambda'} - n_\lambda) \frac{2\Delta_{\alpha 1}}{\omega^2 - \Delta_{\alpha 1}^2} + \sum\limits_{{\substack{\langle \lambda, \lambda'\rangle, \\\lambda = \lambda' }}} F_{\alpha\alpha}^{\lambda\lambda}n_\lambda\frac{\frac{q^2}{m}}{\omega^2 - \left(\frac{q^2}{2m}\right)^2}.
\end{eqnarray}
Thus, to leading order in tunneling, the plasmon gap is approximately given by the form:
\begin{equation}
\omega_{\text{gap}, \alpha}^2\propto {{q_{\text{TF}}d\pi\Delta_{\alpha1}\over m}}.
\end{equation}
as shown in Eq.~(\ref{eq:plasmon_gap-weak_tunneling}). 
Note that the in-phase mode in both the strong and weak tunneling limits is obtained by solving $1-V_{1}(q)\chi^{(0)}_{11}(\boldsymbol{q}, \omega)=0$, and in the long-wavelength limit, we have $\omega_{1}^2(q \to 0) = \frac{2\pi e^2 n_{\text{tot}}}{\kappa m} q$, indicating that the in-phase mode remains unaffected by interlayer tunneling.

\section{Derivation of the plasmon gaps for $N=3$}\label{sec:derivation_N3}

For the $N=3$ case, we have three distinct oscillation modes in the long wavelength limit: the in-phase mode where all layers move together ($\boldsymbol{u}_1 \parallel(1, 1, 1)^T$), one antisymmetric out-of-phase mode where the middle layer remains stationary while the top and bottom layers move in opposite directions ($\boldsymbol{u}_{1} \parallel (1, 0, -1)^T$), and one symmetric out-of-phase mode where the top and bottom layers move together while the middle layer moves in the opposite direction ($\boldsymbol{u}_{3}\parallel (1, -2, 1)^{T}$). In the absence of tunneling, the long-wavelength plasmon modes can be obtained through simple calculations using Eq. (\ref{eq:absence of tunneling}). 

\begin{figure}[htb]
\includegraphics[width=\linewidth]{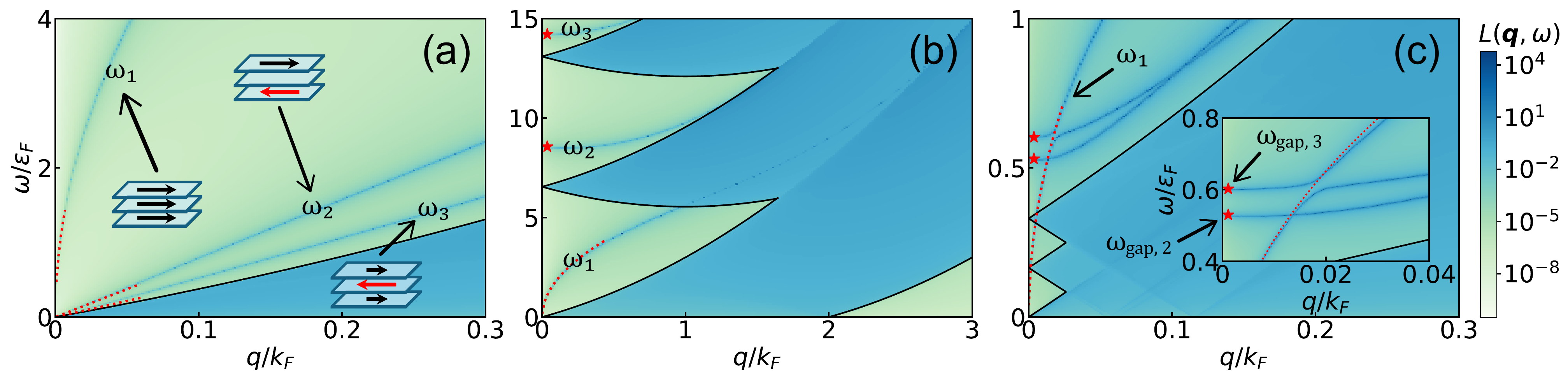}
\caption{
Loss functions of a trilayer system for (a) $n_{\text{tot}}=3\times 10^9$ $\text{cm}^{-2}$ in the absence of tunneling, (b) $n_{\text{tot}}=3\times 10^9$ $\text{cm}^{-2}$ in the presence of tunneling, and (c) $n_{\text{tot}}=3\times 10^{11}$ $\text{cm}^{-2}$ in the presence of tunneling, respectively. For the calculations, the same parameters as in Fig.~\ref{N=4_loss} are used.
}   
\label{N=3_loss}
\end{figure}

In the presence of tunneling, we first consider the out-of-phase mode $\omega_2$. 
Since the asymmetric $\omega_2$ mode is decoupled from the symmetric $\omega_1$ and $\omega_3$ modes, from Eq.~(\ref{long-wavelength limit susceptibility}) we have
\begin{eqnarray}
    1 - V_{2}(q) \sum_{|\lambda-\lambda'| =1}\frac{1}{4}n_{\lambda'}\frac{2\Delta_{\lambda\lambda'}}{\omega^2 - \Delta_{\lambda\lambda'}^2} = 1 - \frac{q_{\rm{TF}}\pi d}{m}\frac{\Delta_{21}}{\omega^2 - \Delta_{21}^2} = 0,
\end{eqnarray}
which implies $\omega_{\rm{gap}, 2}^2 = \Delta_{21}^2 + \frac{q_{\rm{TF}} d \pi}{m}(n_1 - n_2)\Delta_{21}$. Here, we use Eq.~(\ref{eq: U_alpha relation}) for the overlap factors and $\Delta_{21} = \Delta_{32}$.

For the out-of-phase mode $\omega_{3}$, note that the intraband contributions can be neglected in the long-wavelength limit. Then, the $\omega_{\rm{gap}, 3}$ can be obtained as
\begin{equation}
    1 - V_{3}(q)\sum_{|\lambda-\lambda'| = 2}\frac{3}{8}n_{\lambda'}\frac{2\Delta_{31}}{\omega^2 - \Delta_{31}^2} = 1 - \frac{q_{\rm{TF}}\pi d}{2m}\frac{\Delta_{31}}{\omega^2 - \Delta_{31}^2} = 0,
\end{equation}
which implies $\omega_{\rm{gap}, 3}^2 = \Delta_{31}^2 + \frac{q_{\rm{TF}}d\pi}{m}\frac{(n_1 - n_3)}{2}\Delta_{31}$. Thus, the plasmon gaps become
\begin{subequations}
\begin{flalign}
        \omega_{\text{gap}, 2}^2 &= \Delta_{21}^2 + \frac{q_{\text{TF}}d\pi}{m}(n_{1} - n_{3})\Delta_{21},\\
        \omega_{\text{gap}, 3}^2 &= \Delta_{31}^2 + \frac{q_{\text{TF}}d\pi}{m}\frac{(n_1-n_3)}{2}\Delta_{31}.
\end{flalign}
\end{subequations}

For the second mode $\omega_{2}$, only the interactions between symmetric and antisymmetric bands contribute to the oscillation because the corresponding Coulomb mode is antisymmetric. Thus, it is sufficient to consider only the influence of two adjacent bands, and the density of the second band, $n_2$, cancels out since $\Delta_{21}=\Delta_{32}$. 
For the third mode $\omega_{3}$, which is symmetric, only symmetric-symmetric or antisymmetric-antisymmetric bands contribute to the oscillation. Note that near $\Delta_{31}$, unlike the in-phase oscillation, the intraband contribution becomes zero in the long-wavelength limit, and the interband transition between the first and third bands dominates. This mode represents a novel type of a plasmon mode, absent in bilayer systems, and exhibits a larger plasmon gap than that of the second mode. 
Furthermore, since both the in-phase mode $\omega_1$ and the third mode $\omega_3$ are generated by symmetric Coulomb modes, they can interact with one another. The $\sqrt{q}$ dispersion of the in-phase mode allows it to intersect with the dispersions of other symmetric out-of-phase modes in the weak interlayer tunneling regime. In this case, the second mode $\omega_2$, derived from the symmetric-antisymmetric bands, cannot couple with the in-phase mode, whereas the third mode $\omega_3$ can.

\section{Derivation of the plasmon gaps for $N=4$}\label{sec:derivation_N4}

For the $N=4$ case, the out-of-phase modes in the strong coupling limit can be obtained by using Eqs.~(\ref{eq: U_alpha relation}) and (\ref{eq:SM_plasmon gap in the strong-coupling limit}). In the weak coupling limit, considering only the diagonal terms, we obtain
\begin{subequations}\label{eq:N4_weak_coupling}
\begin{flalign}
    1 - V_{2}(q)\chi_{22}^{(0)}(\bm{q}, \omega) &\approx 1 - \left(\frac{C_1}{\omega^2 - \Delta_{21}^2} + \frac{C_2}{\omega^2 - \Delta_{32}^2}\right), \\
    1 - V_3(q) \chi^{(0)}_{33}(\bm{q}, \omega) &\approx 1- \frac{q_{\text{TF}}d\pi}{m}\frac{2(n_1-n_3+n_2-n_4)}{5}\frac{\Delta_{31}}{\omega^2 - \Delta_{31}^2}, \\
    1 - V_{4}(q)\chi^{(0)}_{44}(\bm{q}, \omega) &\approx 1 - \frac{q_{\text{TF}} d \pi}{m}\frac{(5 - \sqrt{5} + \sqrt{10})^2(n_1-n_4)}{100}\frac{2\Delta_{41}}{\omega^2 - \Delta_{41}^2},
\end{flalign}
\end{subequations}
where the constants $C_1$ and $C_2$ are defined as
\begin{subequations}
\begin{eqnarray}
    C_1 &=& \frac{q_{\text{TF}}d\pi}{m} \frac{3+2\sqrt{2}}{5}(n_1-n_2 + n_3 -n_4)\Delta_{21}, \\
    C_2 &=& \frac{q_{\text{TF}}d\pi}{m}\frac{4+\sqrt{2} + \sqrt{5} + \sqrt{10}}{10}(n_2 - n_3)\Delta_{32}.
\end{eqnarray}
\end{subequations}
Here, we retain only the most dominant terms, as the others are negligible. From Eq.~(\ref{eq:N4_weak_coupling}), we can readily determine $\omega_{\text{gap}, 3}$ and $\omega_{\text{gap}, 4}$, as shown in Eq.~(\ref{eq:plasmon_gap-weak_tunneling-N4}). For the $\omega_2$, we need to solve the quadratic equation, yielding
\begin{equation}
    \omega_{\text{gap}, 2\pm}^2 = \frac{\Delta_{21}^2 + \Delta_{32}^2}{2} + \frac{C_1+C_2 \pm \sqrt{(\Delta_{21}^2-\Delta_{32}^2+C_1 - C_2)^2 + 4 C_1 C_2}}{2}.
\end{equation}
Noting that $C_1, C_2 \gg \Delta_{21}, \Delta_{32}$ in the weak coupling limit, we can approximate $\omega_{\text{gap}, 2\pm}^2$ as
\begin{subequations}
\begin{eqnarray}
\omega_{\text{gap}, 2+}^2&\approx& {\Delta_{21}^2+\Delta_{32}^2 \over 2} +C, \\
\omega_{\text{gap}, 2-}^2&\approx& {\Delta_{21}^2+\Delta_{32}^2 \over 2},
\end{eqnarray}
\end{subequations}
where $C= C_1+C_2$, implying that $\omega_{2-}$ remains near the interband transition region.

\section{Derivation of the plasmon gaps for $N\rightarrow \infty$}\label{sec:derivation_N_infinite}

For the $N\rightarrow\infty$ case, we can treat $\theta_{\alpha}$ and $\phi_\lambda$ as continuous variables by replacing $\theta_{\alpha}\rightarrow q_z d$ and $\phi_\lambda \rightarrow k_z d$. Note that $k_z d \in [0, \pi]$ due to Brillouin zone folding. In the absence of tunneling, we only need to use Eq.~(\ref{eq:determinat in the absence of tunneling}) and obtain the following equation:
\begin{equation}
    1-\frac{2\pi e^2}{\kappa q}\frac{1 - e^{-2qd}}{1-2e^{-qd}\cos{(q_zd)} + e^{-2qd}}\chi^{(0)}_{\rm{2D}}(\bm{q}, \omega) = 1-\frac{2\pi e^2}{\kappa q}\frac{\sinh{(qd)}}{\cosh{(qd)} - \cos{(q_z d)}}\chi^{(0)}_{\rm{2D}}(\bm{q}, \omega) = 0.
\end{equation}
From the above equation, we can derive the bulk plasmon dispersions expressed in Eq.~(\ref{eq:plasmon-infinite_N}). 

In the presence of tunneling, we need to calculate the noninteracting density-density response function in Eq.~(\ref{eq:density-density response in presence of tunneling}). When transitioning to a continuous variable, the overlap factor becomes a Dirac-delta function, leaving only the diagonal terms. Consequently, the noninteracting density-density response function with the out-of-plane momentum $q_z$ becomes
\begin{eqnarray}
    \chi_{q_z}^{(0)}(\bm{q}, \omega) &=& g(Nd)^2 \int_0^{\pi/d} \int_0^{\pi/d}\frac{dk_z dk_z'}{\pi^2}\frac{\pi}{2N^2} [\delta(k_z d+k_z' d-q_z d) +\delta(|k_z d-k_z' d|-q_z d) \nonumber\\
    &+& \delta(2\pi - k_zd - k_z'd - q_zd)] \times \int_{}\frac{d^2k}{(2\pi)^2} \frac{f_{\bm{k}, k_z} - f_{\bm{k+q}, k_z'}}{\omega+\varepsilon_{\bm{k}, k_z} - \varepsilon_{\bm{k+q}, k_z'} + i\eta} \nonumber\\
    &=&gd\int_{-\pi/d}^{\pi/d} \frac{dk_z}{2\pi}\int\frac{d^2k}{(2\pi)^2} \frac{f_{\bm{k}, k_z} - f_{\bm{k+q}, k_z+q_z}}{\omega+\varepsilon_{\bm{k}, k_z} - \varepsilon_{\bm{k+q}, k_z+q_z} + i\eta} \nonumber \\
    &=& d \chi^{(0)}_{\rm{3D}}(\bm{q},q_z, \omega),
    \end{eqnarray}
where $\chi^{(0)}_{\rm{3D}}(\bm{q}, q_z, \omega)$ is the noninteracting density-density response function in a 3D anisotropic system. Here, we reformulate the overlap factor as $\delta(k_z'd - k_zd - q_z d)$ by expanding the folded Brillouin zone into $[-\pi/d, \pi/d]$. Similarly, one can obtain the plasmon dispersion. Note that in the long-wavelength limit ($qd, q_zd \ll 1$),
\begin{equation}
    1 - \frac{2\pi e^2}{\kappa q}\frac{d\sinh{(qd)}}{\cosh{(qd)} - \cos{(q_z d)}}\chi^{(0)}_{\rm{3D}}(\bm{q},q_z, \omega) = 1-\frac{4\pi e^2}{\kappa (q^2+q_z^2)} \chi^{(0)}_{\rm{3D}}(\bm{q},q_z, \omega).
\end{equation}
This is consistent with the 3D Lindhard theory with the 3D Coulomb interaction.


\begin{thebibliography}{999}

\bibitem{Pines1966}
D. Pines and P. Nozieres, {\it The Theory of Quantum Liquids} (Benjamin, New York, 1966).

\bibitem{Mahan2000}
G. D. Mahan, \textit{Many-particle Physics}, 3rd ed. (Springer, New York, 2000).

\bibitem{Giuliani2005}
G. F. Giuliani and G. Vignale, \textit{Quantum Theory of the Electron Liquid} (Cambridge University Press, Cambridge, 2005).

\bibitem{Geim2013}
A. K. Geim and I. V. Grigorieva, Van der Waals heterostructures, Nature (London) 499, 419 (2013).

\bibitem{Manfra2014}
M. J. Manfra, Molecular beam epitaxy of ultra-high-quality AlGaAs/GaAs heterostructures: Enabling physics in low-dimensional electronic systems, Annu. Rev. Condens. Matter Phys. \textbf{5}, 347 (2014). 

\bibitem{Polini2020}
M. Polini and A. K. Geim, Viscous electron fluids, Phys. Today \textbf{73}(6), 28 (2020). 

\bibitem{Chung2021}
Y. J. Chung, K. A. Villegas Rosales, K. W. Baldwin, P. T. Madathil, K. W. West, M. Shayegan, and L. N. Pfeiffer, Ultra-high-quality two-dimensional electron systems, Nat. Mater. \textbf{20}, 632 (2021). 

\bibitem{Profumo2010}
R. E. V. Profumo, M. Polini, R. Asgari, R. Fazio, and A. H. MacDonald, Electron-electron interactions in decoupled graphene layers, Phys. Rev. B \textbf{82}, 085443 (2010).

\bibitem{Jang2015}
Y. Jang, E. H. Hwang, A. H. MacDonald, and H. Min,
Stacking dependence of carrier interactions in multilayer graphene systems,
Phys. Rev. B {\bf 92}, 041411(R) (2015).

\bibitem{Fei2015}
Z. Fei, E. G. Iwinski, G. X. Ni, L. M. Zhang, W. Bao, A. S. Rodin, Y. Lee, M. Wagner, M. K. Liu, S. Dai, \textit{et al.}, Tunneling plasmonics in bilayer graphene, Nano Lett. \textbf{15}, 4973 (2015). 

\bibitem{Choi2023}
B. Choi, G. Jeong, S. Ahn, H. Lee, Y. Jang, B. Park, H. A. Bechtel, B. H. Hong, H. Min, and Z. H. Kim, Role of local conductivities in the plasmon reflections at the edges and stacking domain boundaries of trilayer graphene, J. Phys. Chem. Lett. \textbf{14}, 8157 (2023). 

\bibitem{Liu2014}
K. Liu, L. Zhang, T. Cao, C. Jin, D. Qiu, Q. Zhou, A. Zettl, P. Yang, S. G. Louie, and F. Wang, Evolution of interlayer coupling in twisted molybdenum disulfide bilayers, Nat. Commun. \textbf{5}, 4966 (2014). 

\bibitem{Novelli2020}
P. Novelli, I. Torre, F. H. L. Koppens, F. Taddei, and M. Polini, Optical and plasmonic properties of twisted bilayer graphene: Impact of interlayer tunneling asymmetry and ground-state charge inhomogeneity, Phys. Rev. B \textbf{102}, 125403 (2020). 

\bibitem{Zhang2020}
L. Zhang, Z. Zhang, F. Wu, D. Wang, R. Gogna, S. Hou, K. Watanabe, T. Taniguchi, K. Kulkarni, T. Kuo, \textit{et al.}, Moiré lattice-induced formation and tuning of hybrid dipolar excitons in twisted WS$_2$/MoSe$_2$ heterobilayers, Nat. Commun. \textbf{11}, 5888 (2020).

\bibitem{Shin2023}
K. Shin, Y. Jang, J. Shin, J. Jung, and H. Min, Electronic structure of biased alternating-twist multilayer graphene, Phys. Rev. B \textbf{107}, 245139 (2023).

\bibitem{Burg2017}
G. W. Burg, N. Prasad, B. Fallahazad, A. Valsaraj, K. Kim, T. Taniguchi, K. Watanabe, Q. Wang, M. J. Kim, L. F. Register, \textit{et al.}, Coherent interlayer tunneling and negative differential resistance with high current density in double bilayer graphene–WSe$_2$ heterostructures, Nano Lett. \textbf{17}, 3919 (2017). 

\bibitem{Nguyen2019}
P. V. Nguyen, N. C. Teutsch, N. P. Wilson, J. Kahn, X. Xia, A. J. Graham, V. Kandyba, A. Giampietri, A. Barinov, G. C. Constantinescu, \textit{et al.}, Visualizing electrostatic gating effects in two-dimensional heterostructures, Nature \textbf{572}, 220 (2019). 

\bibitem{deVega2017}
S. de Vega and F. J. G. de Abajo, Plasmon generation through electron tunneling in graphene, ACS Photon. \textbf{4}, 2367 (2017).

\bibitem{Alonso-Gonzalez2017}
P. Alonso-González, A. Y. Nikitin, Y. Gao, A. Woessner, M. B. Lundeberg, A. Principi, N. Forcellini, W. Yan, S. Vélez, A. J. Huber, \textit{et al.}, Acoustic terahertz graphene plasmons revealed by photocurrent nanoscopy, Nat. Nanotechnol. \textbf{12}, 31 (2017). 

\bibitem{deVega2019}
S. de Vega and F. J. G. de Abajo, Plasmon generation through electron tunneling in twisted double-layer graphene and metal-insulator-graphene systems, Phys. Rev. B \textbf{99}, 115438 (2019).

\bibitem{Sun2020}
Z. Sun, M. M. Fogler, D. N. Basov, and A. J. Millis, Collective modes and terahertz near-field response of superconductors, Phys. Rev. Res. \textbf{2}, 023413 (2020). 

\bibitem{Menabde2021}
S. G. Menabde, I.-H. Lee, S. Lee, H. Ha, J. T. Heiden, D. Yoo, T.-T. Kim, T. Low, Y. H. Lee, S.-H. Oh, \textit{et al.}, Real-space imaging of acoustic plasmons in large-area graphene grown by chemical vapor deposition, Nat. Commun. \textbf{12}, 938 (2021). 

\bibitem{Goncalves2021}
P. A. D. Gonçalves, T. Christensen, N. M. R. Peres, A.-P. Jauho, I. Epstein, F. H. L. Koppens, M. Soljačić, and N. A. Mortensen, Quantum surface-response of metals revealed by acoustic graphene plasmons, Nat. Commun. \textbf{12}, 3271 (2021). 

\bibitem{Zhu2021}
L. Cui, J. Wang, and M. Sun, Graphene plasmon for optoelectronics, Rev. Phys. \textbf{6}, 100052 (2021).

\bibitem{Huang2022}
T. Huang, X. Tu, C. Shen, B. Zheng, J. Wang, H. Wang, K. Khaliji, S. H. Park, Z. Liu, T. Yang, \textit{et al.}, Observation of chiral and slow plasmons in twisted bilayer graphene, Nature \textbf{605}, 63 (2022). 

\bibitem{Chakraborty2022}
A. Chakraborty, D. Dutta, and A. Agarwal, Tunable interband and intraband plasmons in twisted double bilayer graphene, Phys. Rev. B \textbf{106}, 155422 (2022).

\bibitem{DasSarma1981}
S. Das Sarma and A. Madhukar, Collective modes of spatially separated, two-component, two-dimensional plasma in solids, Phys. Rev. B {\bf 23}, 805 (1981).

\bibitem{DasSarma1998}
S. Das Sarma and E. H. Hwang,
Plasmons in Coupled Bilayer Structures,
Phys. Rev. Lett. {\bf 81}, 4216 (1998).

\bibitem{Borghi2009}
G. Borghi, M. Polini, R. Asgari, and A. H. MacDonald, Dynamical response functions and collective modes of bilayer graphene, Phys. Rev. B \textbf{80}, 241402(R) (2009).

\bibitem{DasSarma2009}
S. Das Sarma and E. H. Hwang, Collective modes of the Massless Dirac Plasma, Phys. Rev. Lett. \textbf{102}, 206412 (2009).

\bibitem{Hwang2009}
E. H. Hwang and S. Das Sarma, Plasmon modes of spatially separated double-layer graphene, Phys. Rev. B \textbf{80}, 205405 (2009).

\bibitem{Gamayun2011}
O. V. Gamayun, Dynamical screening in bilayer graphene, Phys. Rev. B \textbf{84}, 085112 (2011). 

\bibitem{Roldan2013}
R. Roldán and L. Brey, Dielectric screening and plasmons in AA-stacked bilayer graphene, Phys. Rev. B \textbf{88}, 115420 (2013).

\bibitem{Hwang2018}
E. H. Hwang, B. Y.-K. Hu, and S. Das Sarma, Dimensionally mixed coupled collective modes, Phys. Rev. B \textbf{98}, 161304(R) (2018). 

\bibitem{Mohammadi2021}
Y. Mohammadi, Tunable plasmon modes in doped AA-stacked bilayer graphene, Superlattices Microstruct. \textbf{156}, 106935 (2021).

\bibitem{Afanasiev2022}
A. N. Afanasiev, Acoustic plasmons and isotropic short-range interaction in two-component electron liquids, Phys. Rev. B \textbf{106}, 224301 (2022). 

\bibitem{Zhu2013}
J.-J. Zhu, S. M. Badalyan, and F. M. Peeters, Plasmonic excitations in Coulomb-coupled $N$-layer graphene structures, Phys. Rev. B {\bf 87}, 085401 (2013).

\bibitem{Wachsmuth2014}
P. Wachsmuth, R. Hambach, G. Benner, and U. Kaiser,
Plasmon bands in multilayer graphene,
Phys. Rev. B {\bf 90}, 235434 (2014).

\bibitem{Gumbs2016}
G. Gumbs, A. Iurov, J.-Y. Wu, M. F. Lin, and P. Fekete, Plasmon excitations of multi-layer graphene on a conducting substrate, Sci. Rep. \textbf{6}, 21063 (2016).

\bibitem{Lin2018}
C.-Y. Lin, M.-H. Lee, and M.-F. Lin, Coulomb excitations in ABC-stacked trilayer graphene, Phys. Rev. B \textbf{98}, 041408 (2018).

\bibitem{Kim2020}
P. D. T. Kim and M. Nguyen Van, Plasmon modes in $N$-layer graphene structures at zero temperature, J. Low Temp. Phys. \textbf{201}, 311 (2020).

\bibitem{VanMen2021}
N. Van Men, Plasmon modes in $N$-layer silicene structures, 
J. Phys. Condens. Matter {\bf 34}, 085301 (2021).

\bibitem{Cavicchi2024}
L. Cavicchi, I. Torre, P. Jarillo-Herrero, F. H. L. Koppens, and M. Polini, Theory of intrinsic acoustic plasmons in twisted bilayer graphene, Phys. Rev. B \textbf{110}, 045431 (2024). 

\bibitem{Kac1953}
M. Kac, W. Murdock, and G. Szeg\H{o}, On the eigenvalues of certain Hermitian forms, Indiana Univ. Math. J. \textbf{2}, 767 (1953).

\bibitem{Trench2001}
W. F. Trench, \textit{Properties of some generalizations of Kac-Murdock-Szeg\H{o} matrices}, in \textit{Structured Matrices in Mathematics, Computer Science, and Engineering II}, edited by V. Olshevsky (American Mathematical Society, Providence, RI, 2001), Contemp. Math. {\bf 281}, pp. 233–245.


\bibitem{Bogoya2016}
J. M. Bogoya, A. B\"{o}ttcher, S. M. Grudsky, and E. A. Maximenko,
Eigenvectors of Hermitian Toeplitz matrices with smooth simple-loop symbols,
Lin. Alg. Appl. {\bf 493}, 606 (2016).

\bibitem{Fikioris2019}
G. Fikioris, Spectral properties of Kac-Murdock-Szeg\H{o} matrices with a complex parameter, Lin. Alg. Appl. \textbf{553}, 182 (2018). 

\bibitem{Narayan2021}
O. Narayan and B. S. Shastry, Generalized Toeplitz-Hankel matrices and their application to a layered electron gas, J. Phys. A \textbf{54}, 175201 (2021).

\bibitem{Pines1956}
D. Pines, 
Electron Interaction in Solids,
Can. J. Phys. {\bf 34}, 1379 (1956).

\bibitem{Husain2023}
A. A. Husain, E. W. Huang, M. Mitrano, M. S. Rak, S. I. Rubeck, X. Guo, H. Yang, C. Sow, Y. Maeno, B. Uchoa, \textit{et al.},
Pines' demon observed as a 3D acoustic plasmon in Sr$_2$RuO$_4$,
Nature {\bf 621}, 66 (2023).

\bibitem{Zhao2023}
W. Zhao, S. Wang, S. Chen, Z. Zhang, K. Watanabe, T. Taniguchi, A. Zettl, and F. Wang,
Observation of hydrodynamic plasmons and energy waves in graphene,
Nature {\bf 614}, 688 (2023).

\bibitem{Supplemental}
See the Supplemental Material at {\it http://link.aps.org/supplemental/10.1103/wvhd-492f} for details of the Kac-Murdock-Szeg\H{o} Toeplitz matrices, the derivation of the plasmon dispersions for decoupled N-layer systems, the calculation of the noninteracting density-density response function, the derivation of the plasmon dispersions for coupled N-layer systems in the strong and weak tunneling regimes, and the derivation of the plasmon gaps for $N=3$, $N=4$, and $N\rightarrow\infty$.

\bibitem{Fetter1974}
A. L Fetter, Electrodynamics of a layered electron gas. II. Periodic array, Ann. Phys. (N.Y.) \textbf{88}, 1 (1974). 

\bibitem{Jain1985}
J. K. Jain and P. B. Allen, Dielectric response of a semi-infinite layered electron gas and Raman scattering from its bulk and surface plasmons, Phys. Rev. B \textbf{32}, 997 (1985). 

\bibitem{Jain1985a}
J. K. Jain and P. B. Allen, Plasmons in layered films, Phys. Rev. Lett. \textbf{54}, 2437 (1985). 

\bibitem{Christian2022}
C. Boyd, L. Yeo, and P. W. Phillips, Probing the bulk plasmon continuum of layered materials through electron energy loss spectroscopy in a reflection geometry, Phys. Rev. B {\bf 106}, 155152 (2022).

\bibitem{Ahn2021}
S. Ahn and S. Das Sarma,
Theory of anisotropic plasmons,
Phys. Rev. B {\bf 103}, L041303 (2021).




\end{thebibliography}

\begin{thebibliography}{999}
\bibitem{SM_Kac1953}
M. Kac, W. Murdock, and G. Szeg\H{o}, On the eigenvalues of certain Hermitian forms, Indiana Univ. Math. J. \textbf{2}, 767 (1953).

\bibitem{SM_Trench2001}
W. F. Trench, \textit{Properties of some generalizations of Kac-Murdock-Szeg\H{o} matrices}, in \textit{Structured Matrices in Mathematics, Computer Science, and Engineering II}, edited by V. Olshevsky (American Mathematical Society, Providence, RI, 2001), Contemp. Math. {\bf 281}, pp. 233–245.

\bibitem{SM_Bogoya2016}
J. M. Bogoya, A. B\"{o}ttcher, S. M. Grudsky, and E. A. Maximenko,
Eigenvectors of Hermitian Toeplitz matrices with smooth simple-loop symbols,
Lin. Alg. Appl. {\bf 493}, 606 (2016).

\bibitem{SM_Fikioris2019}
G. Fikioris, Spectral properties of Kac-Murdock-Szeg\H{o} matrices with a complex parameter, Lin. Alg. Appl. \textbf{553}, 182 (2018). 

\bibitem{SM_Narayan2021}
O. Narayan and B S. Shastry, Generalized Toeplitz-Hankel matrices and their application to a layered electron gas, J. Phys. A \textbf{54}, 175201 (2021).


\bibitem{SM_Giuliani2005}
G. F. Giuliani and G. Vignale, \textit{Quantum Theory of the Electron Liquid} (Cambridge University Press, Cambridge, 2005).

\end{thebibliography}
\end{document}